\documentclass[aps,prl,twocolumn,groupedaddress,floatfix]{revtex4}
\usepackage{amssymb,amsmath}
\usepackage{graphicx}
\usepackage{subfigure}
\usepackage[english]{babel}
\usepackage{float}
\usepackage{color}

\newcommand{\be}{\begin{equation}}
\newcommand{\ee}{\end{equation}}

\begin{document}

\title{Bounded dynamics of finite PT-symmetric magnetoinductive arrays }

\author{Mario I. Molina}

\affiliation{Departamento de F\'{\i}sica, MSI-Nucleus on Advanced Optics,  and Center for Optics and Photonics (CEFOP), 
Facultad de Ciencias, Universidad de Chile, Santiago, Chile.\\
Tel: 56-2-9787275, Fax: 56-2-2712973, email: mmolina@uchile.cl}


\begin{abstract}
We examine the conditions for the existence of bounded  dynamical phases for finite PT-symmetric arrays of split-ring resonators. The dimer ($N=2$), trimer ($N=3$) and pentamer ($N=5$) cases are solved in closed form, while for $N>5$ results were computed numerically for several gain/loss spatial distributions. It is found that the parameter stability window decreases monotonically with the size of the array. 
\end{abstract}


\maketitle

\clearpage

The study of PT-symmetric systems has attracted a lot of attention during the past few years. In these systems, the effects of loss and gain can balance each other and, as a result,  give rise to a bounded dynamics. These studies are based on the seminal work of Bender and coworkers\cite{bender1, bender2} who showed that non-hermitian Hamiltonians are capable of displaying a purely real eigenvalue spectrum provided the system is symmetric with respect to the combined operations of parity (P) and time-reversal (T) symmetry.
 For one-dimensional systems the PT requirement leads to the condition that the imaginary part of the potential term in the Hamiltonian be an odd function, while the real part be even. The system thus described can experience a spontaneous symmetry breaking from a PT symmetric phase (all eigenvalues real) to a broken phase (at least two complex eigenvalues), as the gain/loss parameter is varied.
To date, numerous PT-symmetric systems have been explored in several fields, from optics\cite{optics1,optics2,optics3,optics4,optics5}, electronic circuits\cite{electronic}, solid-state and atomic physics\cite{solid1,solid2}, to magnetic metamaterials\cite{MM}, among others. The PT symmetry-breaking
phenomenon has been observed in several experiments\cite{optics5,experiments1,experiments2}.

Magnetic metamaterials, on the other hand, consist of artificial structures whose magnetic response can be tailored to a certain extent. A common realization of such system consist of an array of metallic split-ring resonators (SRRs) coupled inductively\cite{SRR1,SRR2,SRR3}. This type of system can feature negative magnetic response in some frequency window, making them attractive for use as a constituent in negative refraction index materials\cite{negative refraction}. A common problem with SRRs is the heavy ohmmic and radiative losses. One possible solution is to endow the SRRs with external gain, such as tunnel (Esaki) diodes\cite{esaki1,esaki2} to compensate for such losses.
 
In the case of some one-dimensional coupled discrete systems, such as a harmonic oscillator array, it has been observed that in the limit of an infinite size array, the system belongs to the broken PT phase, i.e., there are complex eigenvalues making the dynamics unbounded\cite{MM, kevrekidis}. In this work we examine the case of short SRRs arrays, and determine the parameter window inside which the system exhibits a bounded dynamics, and how this window decreases with the size of the system.
 
Let us consider a simple model of a magnetic metamaterial consisting of a finite one-dimensional array of split-ring resonators
including gain/loss terms:
\be
{d^2\over{d \tau^2}} ( q_{n}+\lambda (q_{n+1}+q_{n-1}) )  + \gamma_{n} {d\over{d \tau}} q_{n} + q_{n} = 0\label{eq:1}
\ee
where $\lambda$ is the magnetic interaction coefficient or coupling, and  
$\gamma_{n}$ positive (negative) denotes a ring with loss (gain). In order to satisfy the requirements for PT-symmetry, the spatial distributions of the gain/loss must be odd, $\gamma_{-n} = - \gamma_{n}$. In this work we will focus in binary-like systems with two gain/loss terms and thus examine arrays of the form $\ldots -\gamma_{1}, -\gamma_{2}, -\gamma_{1}, 0, \gamma_{1}, \gamma_{2}, \gamma_{1}, \ldots$, for arrays with an odd number of rings. For arrays with even number of rings the distribution of gain/loss is of the form 
$\ldots -\gamma_{1}, -\gamma_{2}, -\gamma_{1}, \gamma_{1}, \gamma_{2}, \gamma_{1}, \ldots$. This distinction is only meaningful for small arrays and disappears for system of infinite size. Hereafter, and without loss of generality, we will focus on arrays with an odd number of sites (except for the dimer case). Results for the case with an even number of sites are similar.
Since the values of $\gamma_{1}$ and $\gamma_{2}$ are arbitrary, the gain/loss distribution thus introduced allows for many interesting cases to be examined.
In particular we will focus on three cases. The first one is $\gamma_{1}=\gamma$, $\gamma_{2}=-\gamma$, giving rise to the distribution\\
$\ldots \gamma, -\gamma, \gamma, -\gamma, 0, \gamma, -\gamma, \gamma, -\gamma \ldots$.
The second one is $\gamma_{1}=\gamma$ and $\gamma_{2}=0$, which gives rise to the distribution $\ldots 0, -\gamma, 0, -\gamma, 0, \gamma, 0, \gamma, 0, \ldots$. Another interesting case is the one with $\gamma_{1}=\gamma=\gamma_{2}$ that gives  
$\ldots -\gamma, -\gamma, -\gamma, -\gamma, 0, \gamma, \gamma, \gamma, \gamma \ldots$. For this last case, we'll see that in spite of the concentration of loss and gain on opposite sides, the dynamics does possess a stability window for finite arrays lengths.
\vspace{0.5cm}

\begin{figure}[t]
\begin{center}
\noindent
\includegraphics[scale=0.5]{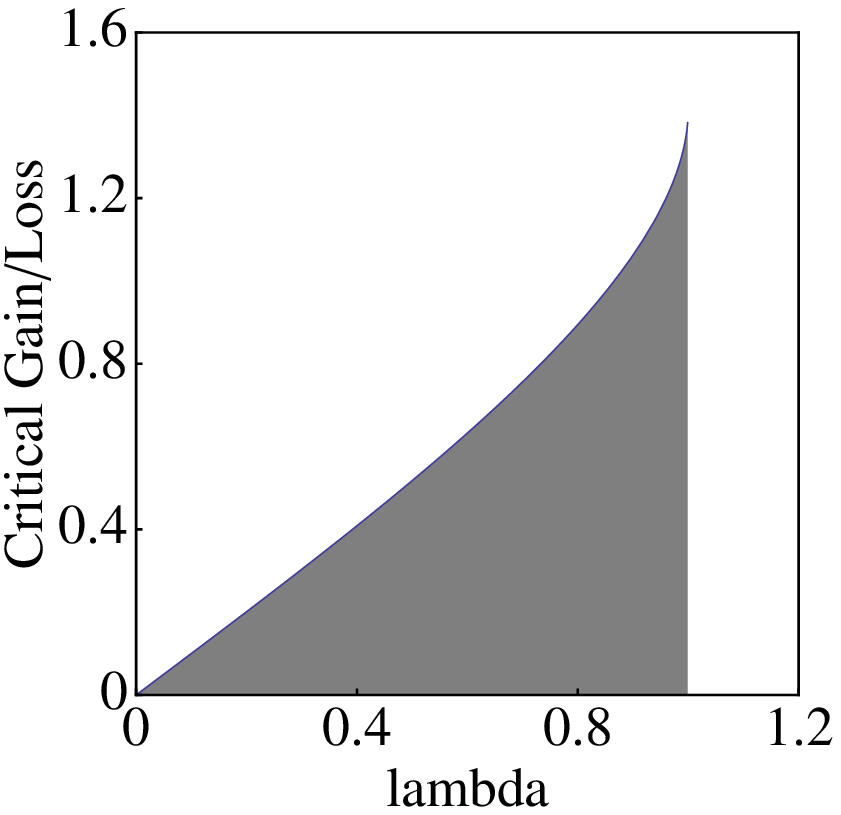}\hspace{0.3cm}
\includegraphics[scale=0.47]{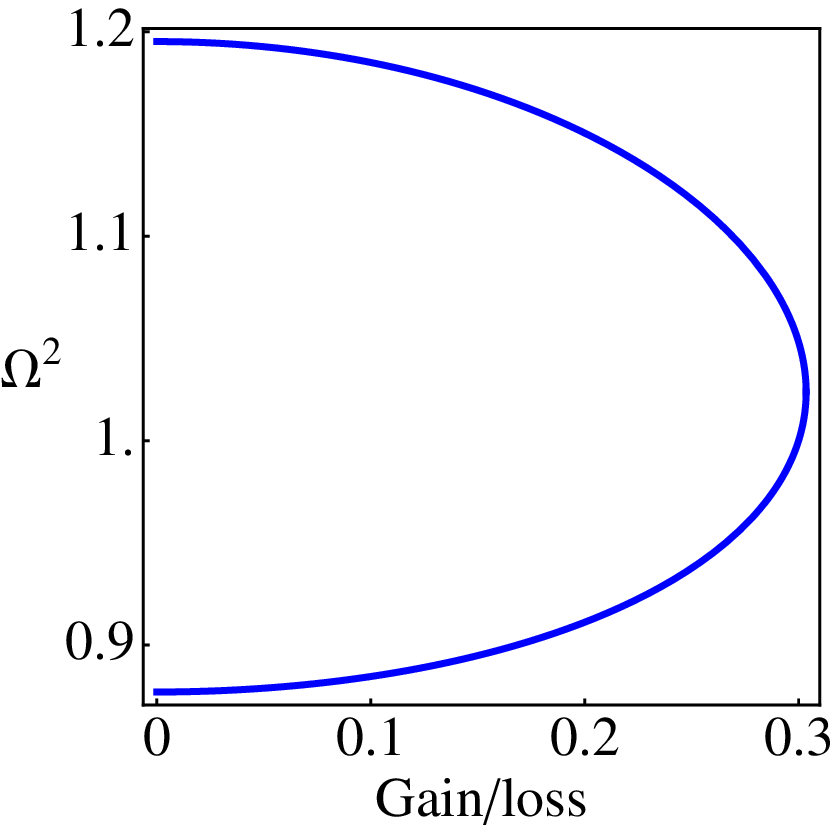}
\caption{Dimer array. Left: Stability region (shaded) in gain/loss-coupling space. Right: Mode frequency squared as a function of the gain/loss parameter for $\lambda=0.3$.}
\label{fig1}
\end{center}
\end{figure}

\noindent
{\em Dimer case ($N=2$)}: In this case, the only possible case is $\gamma, -\gamma$. The dynamical equations read:

\begin{eqnarray}
{d^2\over{d \tau^2}} ( q_{1}+\lambda q_{2})  + \gamma {d\over{d \tau}} q_{1} + q_{1} &=&0\\
{d^2\over{d \tau^2}} ( q_{2}+\lambda q_{1})  - \gamma {d\over{d \tau}} q_{2} + q_{2} &=&0
\end{eqnarray}
We look for stationary modes $q_{1,2}(\tau)=q_{1,2} \exp(i \Omega \tau)$. This leads to the equations,
\begin{eqnarray}
-\Omega^2 (q_{1} + \lambda q_{2}) + i \gamma \Omega q_{1} + q_{1}&=&0\\
-\Omega^2 (q_{2} + \lambda q_{1}) - i \gamma \Omega q_{2} + q_{2}&=&0.
\end{eqnarray}
The condition of the vanishing of the determinant of this linear system leads to a quadratic equation for $\Omega^2$, with solutions:
\be
\Omega = \pm \left[ {2 - \gamma^2 \pm \sqrt{\gamma^4-4 \gamma^2+4 \lambda^2}\over{2(1-\lambda^2)}}\right]^{1/2}.
\ee
We denote the four solutions as $\Omega^{++}, \Omega^{+-}, \Omega^{-+}$ and $\Omega^{--}$. 
The stable phase (unbroken PT symmetry) corresponds to the cases where $\Omega$ is a real quantity. From straightforward examination of Eq.(6), one concludes that the stability window in $\gamma$-$\lambda$ space is given by the area under the curve $\gamma_{c}= \sqrt{2 (1-\sqrt{1-\lambda^2})}$, for $0<\lambda<1$.  Outside $\gamma=\gamma_{c}(\lambda)$, the system  is unstable. Figure \ref{fig1} shows the stability region and also the square frequencies as a function of the gain/loss parameter.
Due to symmetry considerations, only the positive $\gamma$-$\lambda$ sector needs to be explored.

\begin{figure}[t]
\begin{center}
\noindent
\includegraphics[scale=0.75]{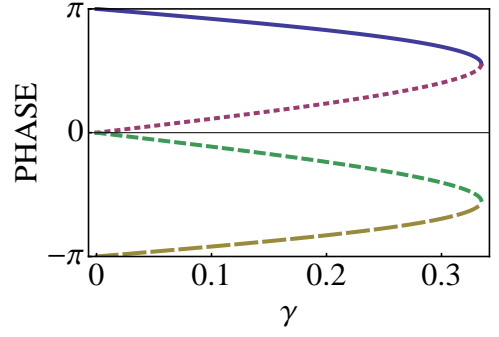}
\caption{Dimer array. Mode phase as a function of the gain/loss parameter.
Solid: $\Omega^{++}$ mode. Dotted: $\Omega^{+-}$ mode. Short dash: $\Omega^{--}$ mode. Long dash: $\Omega^{-+}$ mode. ($\lambda=0.33$)}
\label{fig2}
\end{center}
\end{figure}

From Eq.(4) ( or Eq.(5)) and Eq.(6), it is easy to obtain 
 $|q_{2}/q_{1}| = 1$ and thus, $q_{1}$ and $q_{2}$ differ by a phase only. We have four branches for the phase, corresponding to each of the four solutions. Figure \ref{fig2} shows the phase of all solutions as a function of the gain/loss parameter. 

\vspace{0.5cm}

\noindent
{\em Trimer case ($N=3$)}: Here the gain/loss distribution has the form $-\gamma, 0, \gamma$. The stationary state equations have the form
\begin{eqnarray}
-\Omega^2 (q_{1} + \lambda q_{2}) + i \gamma \Omega q_{1} + q_{1}&=&0\\
-\Omega^2 (q_{2} + \lambda (q_{2} + q_{3})) + q_{2} &=&0\\
-\Omega^2 (q_{3} + \lambda q_{2}) - i \gamma \Omega q_{3} + q_{3}&=&0.
\end{eqnarray}
with eigenvalues
\begin{eqnarray}
\Omega & = & \pm 1\\
\Omega & = & \pm \left[ {{2 - \gamma^2 \pm \sqrt{-4 \gamma^2+\gamma^4+8 \lambda^2}}\over{2-4 \lambda^2}} \right]^{1/2}
\end{eqnarray}
The condition that {\em all} $\Omega$ be real leads to the conditions $\lambda<1/\sqrt{2}$ and $\gamma<\gamma_{c}=\sqrt{2-\sqrt{4-8 \lambda^2}}$.
\begin{figure}[b]
\begin{center}
\noindent
\includegraphics[scale=0.5]{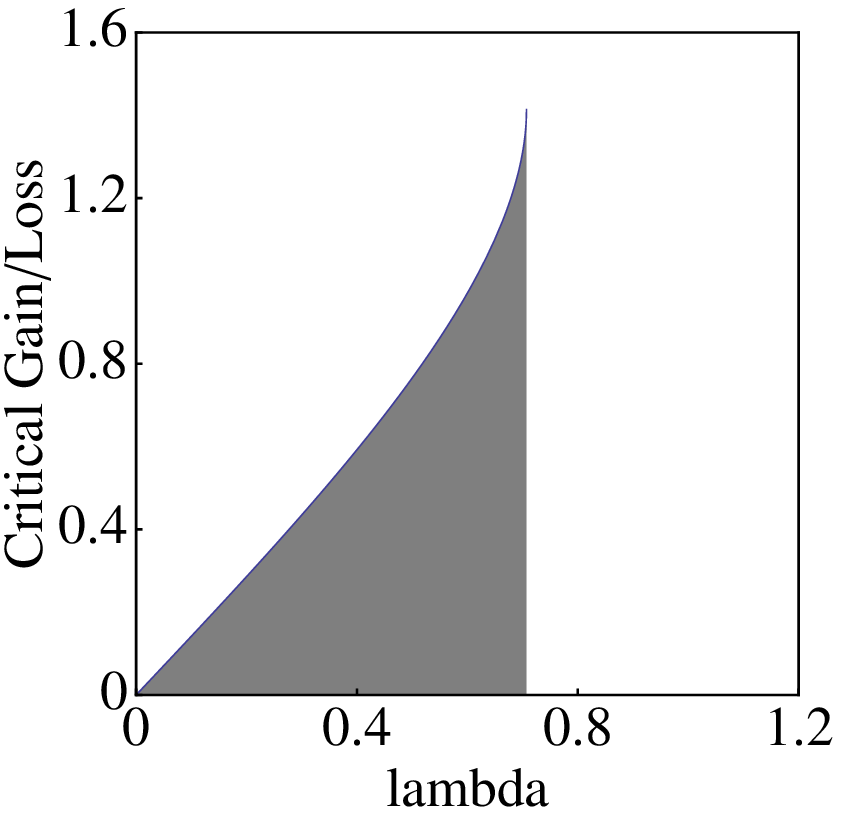}\hspace{0.3cm}
\includegraphics[scale=0.47]{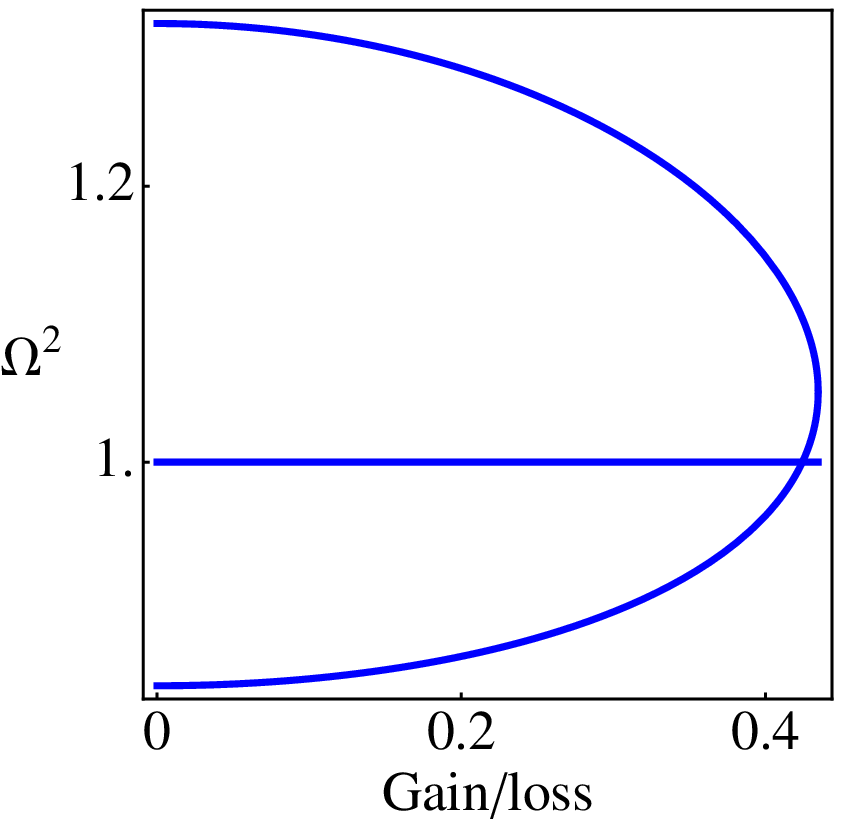}
\caption{Trimer array. Left: Stability region (shaded) in gain/loss-coupling space. Right: Mode frequency squared as a function of the gain/loss parameter, for $\lambda=0.3$.}
\label{trimer}
\end{center}
\end{figure}
Figure \ref{trimer} shows the window of real eigenvalues in gain/loss-coupling space. It is qualitatively similar to the dimer case, but the allowed coupling interval is smaller. The figure also shows the square of the real frequency as a function of gain/loss, for a given coupling value.

\noindent
{\em Pentamer case ($N=5$)}: Here the gain/loss distribution can have three possible forms: $\gamma,-\gamma, 0, \gamma, -\gamma$, or $0,-\gamma, 0, \gamma, 0$, or $-\gamma,-\gamma, 0, \gamma, +\gamma$. We will focus on the first case since is more amenable to an exact form solution (Numerical results show that the other two cases display similar behavior). The stationary state equations have the form
\begin{eqnarray}
-\Omega^2 (q_{1} + \lambda q_{2}) + i \gamma \Omega q_{1} + q_{1}&=&0\\
-\Omega^2 (q_{2} + \lambda (q_{1} + q_{3})) - i \gamma \Omega q_{2} + q_{2} &=&0\\
-\Omega^2 (q_{3} + \lambda(q_{2} + q_{4})) + q_{3}&=&0\\
-\Omega^2 ( q_{4}+\lambda (q_{3}+q_{5}) ) + i \gamma \Omega q_{4} + q_{4}&=&0\\
-\Omega^2 ( q_{5}+\lambda q_{4} ) - i \gamma \Omega q_{5} + q_{5} &=&0.
\end{eqnarray}
with eigenvalues
\begin{eqnarray}
\Omega & = & \pm 1\\
\Omega & = & \pm \left[ {{2 - \gamma^2 \pm \sqrt{-4 \gamma^2+\gamma^4 + 4 \lambda^2}}\over{2-2 \lambda^2}} \right]^{1/2}\\
\Omega & = & \pm \left[ {{2 - \gamma^2 \pm \sqrt{-4 \gamma^2+\gamma^4 + 12 \lambda^2}}\over{2-6 \lambda^2}} \right]^{1/2}
\end{eqnarray}
The condition that {\em all} $\Omega$ be real leads to the conditions $\lambda<1/\sqrt{3}$ and $\gamma<\gamma_{c}=\sqrt{2(1-\sqrt{1-\lambda^2)}}$.
\begin{figure}[t]
\begin{center}
\noindent
\includegraphics[scale=0.5]{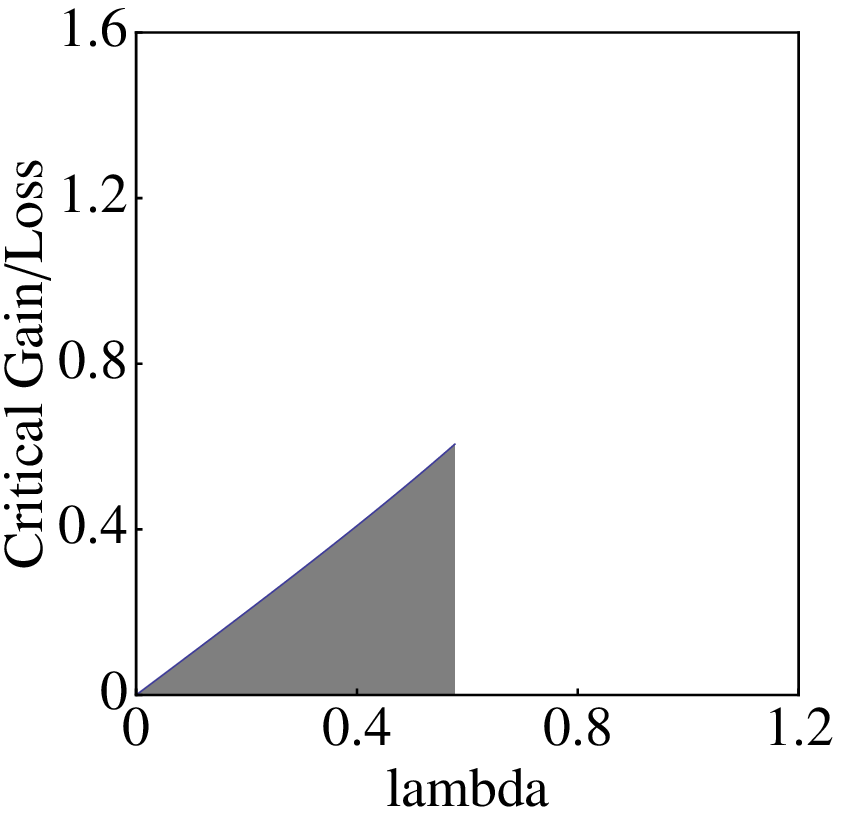}\hspace{0.3cm}
\includegraphics[scale=0.47]{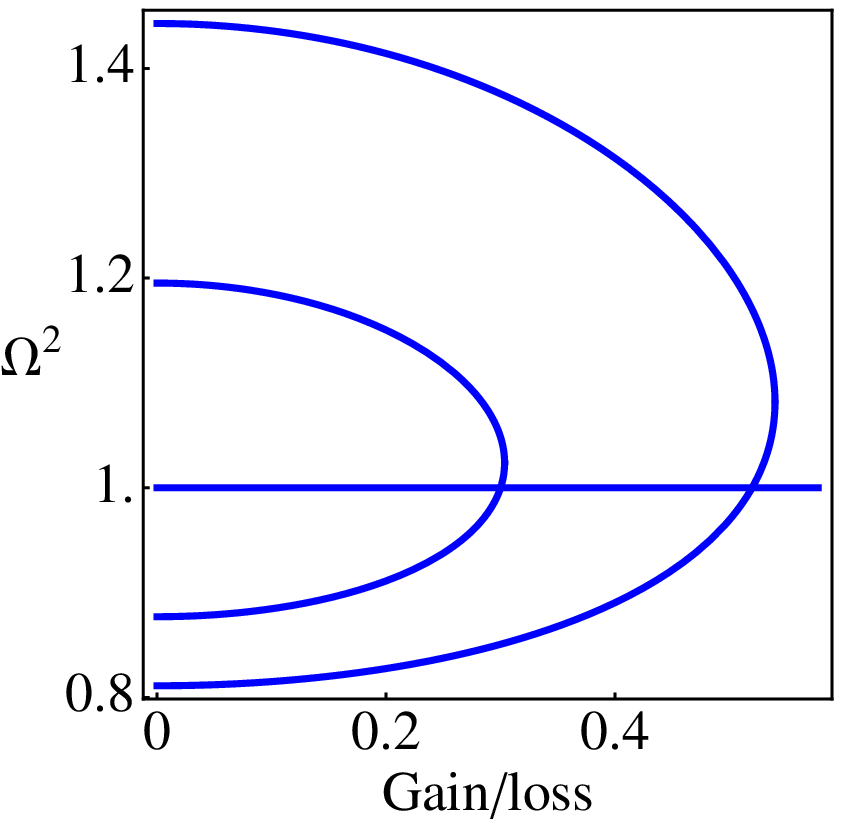}
\caption{Pentamer array. Left: Stability region (shaded) in gain/loss-coupling space. Right: Mode frequency squared as a function of the gain/loss parameter, for $\lambda=0.3$.}
\label{pentamer}
\end{center}
\end{figure}
Figure \ref{pentamer} shows the stability window in gain/loss-coupling space, as well as the square frequency as a function of gain/loss, for a fixed coupling value. As we can see, the stability window is substantially smaller than the one for the dimer and trimer cases.

\vspace{0.5cm}

\noindent
{\em Short chains ($N>5$)}:  Let us consider now the case of finite arrays, where the stationary-state equations are given by
\be
-\Omega^2 [ q_{n} + \lambda (q_{n+1}+q_{n-1}) ] + (1+i \Omega \gamma_{n}) q_{n} = 0.\label{eq:7}.
\ee 
The  $\{\gamma_{n}\}$ distribution we consider has the general form: $\ldots -\gamma_{1}, -\gamma_{2}, -\gamma_{1}, 0, \gamma_{1}, \gamma_{2}, \gamma_{1}, \ldots$. 
We will compute  the relevant eigenvalues numerically from the vanishing of the determinant of linear system (\ref{eq:7}) and will  focus on the parameter values in gain/loss - coupling space where the eigenvalues are purely real, thus denoting a bounded dynamical regime.

\begin{figure}[t]
\begin{center}
\noindent
\includegraphics[scale=0.65]{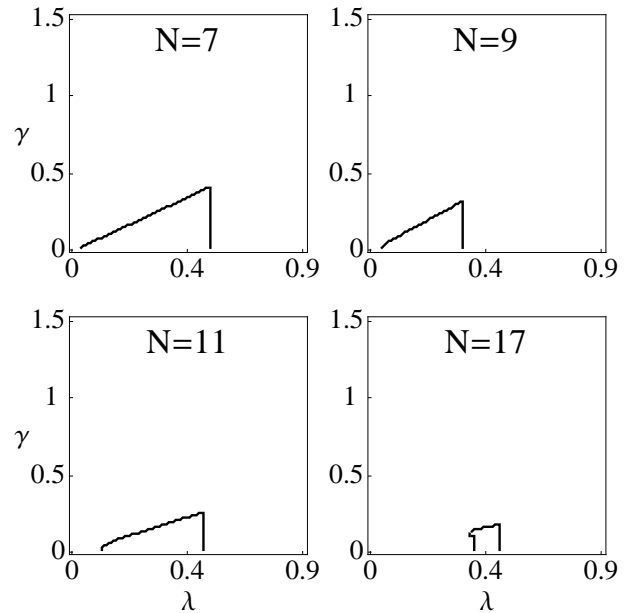}
\caption{Stability regions (area under curve) in gain/loss-coupling space for several array lengths, for the case $\ldots, -\gamma, +\gamma, -\gamma,\ \text{0}\ , +\gamma, -\gamma, +\gamma,\dots$.}
\label{fig3}
\end{center}
\end{figure}

\noindent
{\em Case a}.\ \ We start with the case
$\gamma_{1}=-\gamma_{2}\equiv \gamma$ that gives rise to the distribution 
$\ldots, -\gamma, +\gamma, -\gamma, \text{O}, +\gamma, -\gamma, +\gamma,\dots$. Some results are shown in Fig.\ref{fig3}. It is clear that, as the size of the array increases, the stability region shrinks, and disappears 
altogether around $N=20$. This is consistent with a previous result\cite{MM} stating that in the infinity size limit, the dynamics is always unstable (that is, the system belongs to the broken PT phase).\\
\noindent
{\em Case b}. Now we take $\gamma_{2}\rightarrow 0$ and $\gamma_{1}\equiv \gamma$, that is, the distribution 
$\dots, 0, -\gamma,  0, -\gamma, \text{O}, +\gamma, 0, +\gamma, 0, +\gamma,0,\dots$. Notice how the gain and loss portions are now separated and on each side they are rather diluted. The stability phase diagrams for this case (not shown) are qualitatively similar to the previous case, although, for a given array size, the stability windows are smaller.

\noindent
{\em Case c}.\ \ Now we take $\gamma_{2}=\gamma_{1}\equiv \gamma$. The gain/loss distribution is\\
$\ldots, -\gamma, -\gamma, -\gamma, -\gamma,\text{O}, +\gamma, +\gamma, +\gamma, +\gamma, \ldots  $
Now the gain and loss on each side are densely populated and the resulting area of the stability windows (not shown) while qualitatively similar to the previous cases, drop even faster.

\begin{figure}[t]
\begin{center}
\noindent
\includegraphics[scale=0.65]{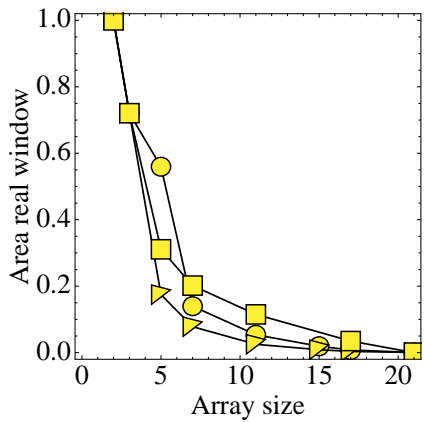}
\caption{Normalized size of area in gain/loss-coupling space with purely real eigenvalues as a function of array size for the gain/loss distributions  ``a'' (squares), ``b'' (circles), and ``c'' (triangles).  }
\label{fig4}
\end{center}
\end{figure}

Figure \ref{fig4} shows a summary of the results obtained for
the size of the stability window as a function of the array
length, for the three cases considered. We clearly see that
the stability region decreases rather abruptly with $N$,
and that for a given $N$, we have $\mbox{area}\ a < \mbox{area}\ b < \mbox{area}\ c$,
for $N > 5$.
\newpage

\noindent
{\em Conclusions}. We have examined the dynamics of finite, one-dimensional PT-symmetric arrays of split-ring
resonators, that constitute the simplest model of a magnetic metamaterial, and have focussed on the conditions in parameter 
space where this dynamics is bounded, i.e., all eigenvalues 
are real. The dimer, trimer and pentamer cases were solved in closed form,
while for larger but finite arrays, results were obtained
numerically. It was found that in general, the stability 
region decreases abruptly with an increase in system size.
This tendency seems generic for gain/loss distributions
based on two gain/loss parameters, and is in agreement
with recent results for 
infinite discrete arrays\cite{MM,kevrekidis}.

This work was supported in part by Fondo Nacional de Ciencia y Tecnolog\'{\i}a (Grant 1120123), Programa Iniciativa Cient\'{\i}fica Milenio (Grant P10-030-F), and Programa de Financiamiento Basal (Grant FB0824/2008).

\end{document}